# O/IR Polarimetry for the 2010 Decade (GAN): Science at the Edge, Sharp Tools for All

A Science White Paper for the:
**The Galactic Neighborhood (GAN) Science Frontiers Panel** of the
Astro2010 Decadal Survey Committee

Lead Authors:

Dan Clemens
Astronomy Department
725 Commonwealth Ave
Boston University
Boston, MA 02215
(617) 353 – 6140 (ph)
clemens@bu.edu

B-G Andersson
Stratospheric Observatory
   for Infrared Astronomy
NASA Ames Res. Center
Mail Stop 211-3
Moffett Field, CA 94035
(650) 604 6221 (ph)
bg@sofia.usra.edu

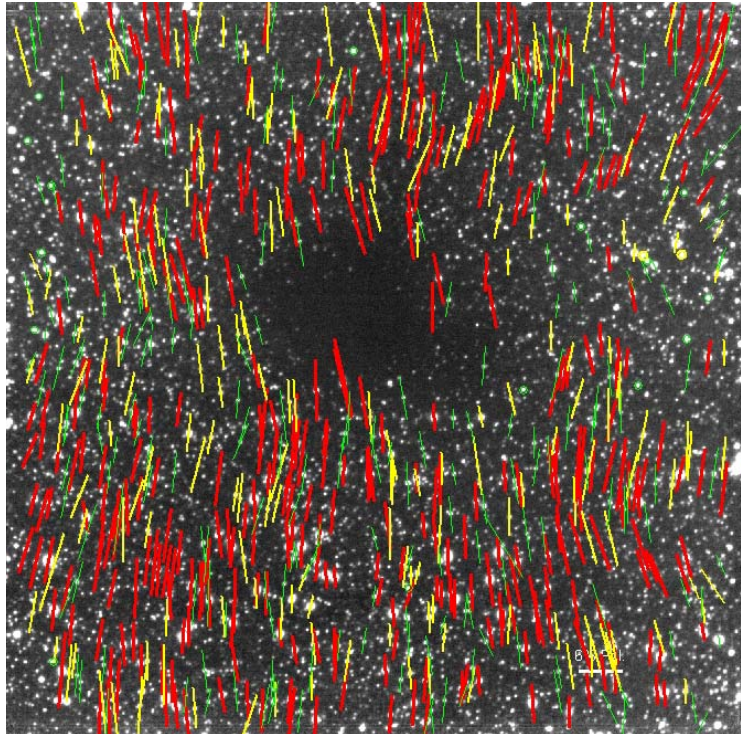

*NIR imaging polarimetry of the FeSt globule in the Pipe Nebula (Clemens et al. 2009, in prep.)*

<u>Contributors and Signatories</u>

| | |
|---|---|
| Andy Adamson | UKIRT, JAC, Hilo |
| David Axon | Rochester Institute of Technology |
| James De Buizer | SOFIA, NASA Ames |
| Alberto Cellino | Osservatorio Astronomico di Torino, Italy |
| Dean C. Hines | Space Science Institute, Corrales, NM |
| Jennifer L. Hoffman | University of Denver |
| Terry Jay Jones | University of Minnesota |
| Alexander Lazarian | University of Wisconsin |
| Antonio Mario Magalhaes | University of Sao Paulo, Brazil |
| Joseph Masiero | University of Hawaii |
| Chris Packham | University of Florida |
| Marshall Perrin | UCLA |
| Claudia Vilega Rodrigues | Inst. Nac. De Pesquisas Espaciais, Brazil |
| Hiroko Shinnaga | CalTech |
| William Sparks | STScI |
| John Vaillancourt | CalTech |
| Doug Whittet | RPI |

*Overview and Context: Polarimetry as a cross-cutting enterprise*

Photometry, spectroscopy, and polarimetry together comprise the basic toolbox astronomers use to discover the nature of the universe. Polarimetry established the Unified Model of AGN and continues to yield unique and powerful insight into complex phenomena. Polarimetry reveals the elusive magnetic field in the Milky Way and external galaxies, allows mapping of features of unresolved stars and supernovae, uncovers nearby exoplanets and faint circumstellar disks, and probes acoustic oscillations of the early universe.

Polarimetry is practiced across the full range of accessible wavelengths, from long wavelength radio through gamma rays, to provide windows into phenomena not open to photometry, spectroscopy, or their time-resolved variants. At some wavelengths, the U.S. leads the world in polarimetric capabilities and investigations, including ground-based radio, through the VLA and VLBA. At other wavelengths, the U.S. is currently competitive: in submm the CSO and the JCMT have historically pursued similar science problems.

In ground-based O/IR, the situation is considerably worse, with no optical or NIR polarimeters available on Gemini (Michelle is MIR, only) or any NOAO-accessed 4 m telescope, as the table below shows. Over the past decade and more, Canadian and European astronomers have enjoyed unique access to state-of-the-art polarimeters and have used this access to vault far past the U.S. in many science areas.

| Telescope | Aperture | Instrument | Waveband | Polar. Mode | U.S. Access ? |
|---|---|---|---|---|---|
| IRSF (SAAO) | 1.4m | SIRPOL | NIR | Imaging | No |
| Perkins (Lowell) | 1.8m | Mimir, PRISM | NIR, Optical | Imaging | Private |
| HST | 2.4m | WFPC2, ACS, NICMOS | Optical, Optical, NIR | Imaging | Yes |
| Nordic Optical | 2.5m | TURPOL | Optical | Photopol | No |
| MMT | 6.5m | MMTPOL | NIR | Imaging | Private |
| LBT | 2x8.4m | PEPSI | Optical | Spectropol | Private |
| Gemini | 8m | Michelle | MIR | Imaging | Yes |
| Keck | 10m | LRIS | Optical | Spectropol | Private |
| GTC | 10m | CanariCam | MIR | Imaging | No |

In space, NICMOS, ACS, and WFPC-2[1] on HST have permitted imaging polarimetry down to 0.1% precision, and may represent the most general purpose O/IR access for U.S. astronomers. Neither the Spitzer Space Telescope nor JWST provides, or will provide, any polarimetric capability.

The dwindling U.S. access to this crucial third leg of the light analysis tripod has also become self-fulfilling, as students receive little exposure to polarimetric techniques and scientific advances as the number of practitioners able to teach students declines.

Nevertheless, polarimetric studies in O/IR have already revealed a great deal about star and planet formation processes, stars and their evolution, the structure of the Milky Way, and the nature and origin of galaxies and their active nuclei – details which cannot be discovered using pure photometric or spectroscopic methods. For example, NIR imaging polarimetric studies by the SIRPOL group (e.g., Tamura et al. 2007) reveal the details of the magnetic fields lacing nearby, star-forming molecular cloud cores and the embedded reflection nebulae and disks associated with their newly formed stars. Further, the race to find and image exoplanets will use polarimetry internal to the two extreme adaptive optics coronagraphs now under construction: SPHERE/ZIMPOL (for the VLT: Joos 2007) and GPI (for Gemini: Macintosh et al. 2006).

---

[1] NICMOS and much of ACS are currently off-line until Servicing Mission 4 (SM4); WFPC-3 will replace WFPC-2 but will have no polarimetric capability.

Theoretical efforts have recently advanced our understanding of the origin of dust grain alignment by demolishing old theory and offering new, testable predictions. The new paradigm of radiative aligned torques (e.g., Lazarian & Hoang 2007; Hoang & Lazarian 2008) is removing old doubts concerning magnetic alignment while leaving a wide parameter space open for observational testing and future theoretical refinement.

Polarimetric modeling has also entered a modern age, one characterized by huge model grids, spectacular dynamic ranges, and closer coupling to observational measurements. The active debate on the relative importance of magnetohydrodynamics (e.g., Li et al. 2004) vs. pure hydrodynamics (e.g., Padoan & Nordlund 2002) is shining new light into the nature of magnetic fields in the ISM and in star formation. Meanwhile, radiative transfer modeling guides detailed interpretation of complex polarized line profiles from aspherical stellar winds (e.g., Harries 2000) and supernovae (e.g., Hoeflich 2005; Kasen et al. 2006; Hoffman 2007). This upcoming decade is sure to see careful, detailed comparisons between high-resolution model simulations and observational data that will lead to sharp new insights into many astrophysical scenarios.

The promise evident in the new, niche polarimetric instruments and the surveys they will perform will drive cutting-edge science in the upcoming decade. Yet finding answers for many key questions requires open community access to general purpose, precision polarimeters on large telescopes, as well as opportunities for student training.

*Example Polarimetry Science Areas for the next decade*

Within the broad area covered by The Galactic Neighborhood panel, polarimetric studies for the upcoming decade will address several key questions: (1) What role(s) does the magnetic field play in assembling diffuse interstellar material into new atomic and molecular clouds? Which field properties (strength, orientation) decide the fate of star-forming and quiescent cloud cores? (2) Which physical mechanism most aligns dust grains? (3) How are interstellar grains and grain mantles structured? (4) Will the new theory of atomic magnetic realignment (AMR) launch a viable new probe of cosmic magnetic fields? These questions are fundamental to understanding the nature of the Milky Way, yet can only be answered through new polarimetric observations, theory, and open access to precision polarimeters on large telescopes. Examples of open challenges ripe for polarimetric progress for the upcoming decade include:

- Resolve the 3D structure of the Galactic magnetic field

    In the diffuse, ionized ISM the magnetic field is traced over kpc scales via radio Faraday rotation (e.g., Brown et al. 2007). The observations reveal asymmetries in the field about the midplane (Han et al. 2006), and spiral arm field reversals. In the neutral ISM, radio Zeeman effect measurements and background star polarimetry (see Heiles 2000), tracing magnetically-aligned dust grains, are the chief probes. However, large areas of the neutral medium have eluded both techniques, due to low pulsar area density and deep dust extinction in the midplane. Infrared stellar polarimeters are needed to survey substantial parts of the Galactic disk and to probe to the multi-kpc depths necessary to reveal the 3D distribution of the field. This will advance both magnetic field studies and our understanding of how the Milky Way was assembled. Further, models of the origin and maintenance of the Galactic field (e.g., Gnedin et al. 2000; Zweibel 2003) will be accessible to quantitative testing through combining new optical polarimetry of the solar neighborhood (e.g. Alves et al. 2008), the IR-revealed 3D nature of the

field, and the relation of near and distant magnetic structures to superbubbles (including the Local Bubble; Leroy 1999) and HII regions.

The magnetic field in the diffuse, ionized ISM is in rough equiparition with all other energy densities. Self-gravity, however, becomes important for atomic clouds and may be the dominant energy density for molecular clouds. Yet out of the just finished decade, a new, complex, and highly dynamic ISM has emerged (e.g., the $^{13}$CO Galactic Ring Survey, Jackson et al. 2006; the VLA Galactic Plane Survey, Stil et al. 2006; WHAM, Reynolds 2004; and GLIMPSE, Benjamin et al. 2003) with likely roles for the magnetic field on all scales. The ubiquitous O VI in the Galactic disk, seen with FUSE (Bowen et al. 2008), reveals long-lived interfaces between warm/cool and coronal gas. Do magnetic fields extend these interface lifetimes by restricting electron transport across field lines? Is the creation of interstellar clouds dominated by conditions in the magnetic field, or by turbulent motions, as the observations of Crutcher et al. (2008) imply? To answer these questions, the magnetic field must be revealed in great detail and be traced across the full range of environments, from diffuse ionized to cool atomic to dense molecular. This requires sensitive polarimetric observations across radio, FIR/submm, and O/IR with coordinated targets, campaigns, and surveys. The requisite instruments and telescopes are either just now becoming available, or should be encouraged to be so, in order for these key science goals to be met in this decade.

- Resolve the magnetic grain alignment mechanism.

The major unknown for magnetic interpretation of ISM polarimetry, both for field topology and field strength (see "*Exceptional Discovery*" section, below) - is the physics of dust grain alignment. ISM polarization **is** due to magnetically aligned, asymmetric dust grains, yet only recently has a quantitative, testable theory emerged. This theory (Lazarian & Hoang 2007a, b, and references therein) and new observations (Andersson & Potter 2007; Whittet et al. 2008) now provide strong support for radiatively driven alignment. However, detailed understanding of the alignment mechanism is needed for accurate interpretation of ISM polarization observations. Radiative grain alignment theory has dependences on grain size, radiation wavelength, the angle between the radiation and magnetic field, and possibly the rate of molecular hydrogen formation (Hoang & Lazarian 2008). New observations, over a wide range of environments, are needed to test the underlying grain alignment physics and to provide reliable tracers of the magnetic field from observed polarimetry. Secondary mechanisms depend on the detailed characteristics of the grains (Draine & Lazarian 1998); improved understanding of radiative alignment will also probe related grain micro-physics. Comprehensive tests of radiative alignment theory require that the observations utilize large samples, going to deep extinctions - both requiring large telescopes and powerful precision polarimeters from the U-band through the MIR.

- Strip bare, interstellar grains and grain mantles, to peer into dark places.

What are dust grains made of, what are their structures, and how do they grow? Spectroscopy of solid-state features, including ice features from grain mantles, provides tremendous information on structure and composition of the grains as well as on their environments. For example, Whittet et al. (1989) showed ice mantles sublimate as grains are heated by the diffuse radiation fields penetrating dark clouds, requiring a threshold extinction into the cloud, of about $A_V$=3 for water and about $A_V$≈5 for CO, for ices to form. Yet **spectropolarimetry** of many of the same feature reveals much about the grains that is missed otherwise. Hough et al. (2008) used NIR spectropolarimetry to compare the ice absorption and ice polarization spectral profiles to constrain the axis ratio of the grains and relative volumes of grain cores to ice mantles for the line of sight to Elias 16 in Taurus. The powerful combination of spectroscopy and polarimetry of solid state features is a new tool that, if encouraged for the upcoming decade, will probe grain properties across a full range of environments, timescales, and illuminations.

The polarization of these grain mantle ices also provides a powerful new tool to reveal *in situ* magnetic fields deep in dark clouds. The ices exist only over a fairly narrow range of conditions (for CO ice, $T_{dust} < 17K$), so this "ice polarization" provides the best connection between the FIR/submm polarimetry of only the densest parts of clouds and the optical and NIR polarimetry, which does best in cloud peripheries. Indeed, Hough et al. see polarization position angles for the continuum, water ice, and CO ice that systematically shift, revealing magnetic field rotation with depth into the cloud. This new tool for "magnetic tomography" of molecular clouds should be fully exploited in the upcoming decade to measure the size scales of ambipolar diffusion, magnetic pinch topology, and how these relate to star formation or accretion rates.

- Atomic Magnetic Realignment (AMR): a New Magnetic Field Diagnostic

Yan and Lazarian (2008, and ref.s therein) have predicted a new effect, "atomic magnetic realignment" (AMR), in which the magnetic field geometry in the diffuse interstellar medium, or in circumstellar matter, is determined from the linear polarization of absorption or fluorescence emission lines pumped by an anisotropic radiation field. AMR is potentially more powerful than radio Zeeman measurements, since it is sensitive to weaker fields and works in hot gas. AMR may also be more powerful than dust alignment and synchrotron methods, since it is sensitive to 3D geometry, gas properties and velocity. The effect has been observed in the Sun (Landi degl'Innocenti 1999) but not yet outside the solar system, because of the lack of suitable instrumentation. The emission line effect would be observable using moderate resolution spectropolarimeters that are sensitive to faint diffuse line emission (e.g., Nordsieck 2008). The huge potential AMR offers magnetic field science requires swift action to test for, and hopefully exploit, this wonderful new technique.

*Exceptional Discovery Potential Area: Widespread Precision **B** Strengths*

Magnetic field strength is a crucial factor for the nature and structure of the ISM. Strong fields can quench gas dynamic motions, or direct flows along field lines. Weak fields may still regulate star formation (e.g., Duffin & Pudritz 2008) and models that include magnetic fields (e.g., Galli et al. 2001; Tassis & Mouschovias 2007; Kudoh et al. 2007; Elmegreen 2007) find that the ratio of field strength to gravitational energy is a chief outcome determinant. But until now, routine, high-quality measurements of field strengths in the astrophysical settings of most interest have been difficult to impossible.

This upcoming decade will see a major change, resulting from new and improved techniques, instrumentation, theory and interpretation. Ready access to magnetic field strengths for large numbers of clouds and cores will make it easy to include magnetic strengths in ISM studies and permit direct tests of modern theories.

The observationally simplest method for measuring the field strength is the CF technique (Chandrasekhar & Fermi 1953), which probes the balance between magnetic field strength and turbulent gas motions, using the dispersion in the polarization angles and the velocity profile width of relevant gas tracer (H I, CH, K I, Ca II, OH, CO, etc.).

The CF technique is based on two concepts: (1) the MHD equations of motion give a one-dimensional, negative pressure directed along the field lines (equivalent to a tension in the field lines); and (2) the vibration amplitude of a string is determined by its tension, its mass, and the driving force. In the past, projection effects and other uncertainties appeared to limit the quality of the derived strengths, but recent theory (Falceta-Gonçalves et al. 2008), and observations, argue otherwise. While changes in the average field along the line of sight affect the observed polarization dispersion (Ostriker et al. 2001; Andersson & Potter 2005), densely sampled polarization maps (e.g., Alves et al. 2008), preferably with stellar distance estimates, can overcome these complications. New wide-area and NIR imaging polarimeters (e.g., Clemens et al. 2007), are needed to provide the necessary sampling density for full exploitation of the CF technique and so to enable a new era of magnetic field strength determinations.

*What is Needed to Meet the Science Goals within the Decade*

What key observations, theory, and instrumentation are needed to achieve the science goals within the next decade? Our evaluation of the upcoming opportunities, challenges, and technical readiness leads to the following recommendations:

1. *Build precision polarimetric capability into new O/IR instruments for large telescopes and space missions. Design polarimetry in from the beginning, not as "add-ons"*.
   - New "niche" instruments, such as GPI, SPHERE/ZIMPOL, and HiCIAO on Subaru rightly exploit polarimetry to meet their exoplanet and circumstellar disk objectives, but general purpose instruments with polarimetric capability are lacking at virtually all large, open-access US telescopes - ground-based, airborne, and space-based, especially in the infrared.
   - Key science questions cannot be answered unless US astronomers have access to precision (photon-noise limited) polarimetric capability. To retain precision capability, polarimetric capabilities must be a considered in the initial design of the instruments, not as a later "add-on".
   - Exoplanet characterizations may favor bluer wavelengths, to enhance detection of Rayleigh scattering, but fractional polarizations are low, requiring access to the largest telescopes to enable sampling adequate volumes.

2. *Encourage polarimetric surveys with LSST*.
   - So much of the sky has never been explored polarimetrically that it represents a "new frontier" for optical wavelengths. Many key science advances are in the Polarimetry White Paper to the LSST Consortium[2].
   - LSST surveys offer one of the best chances of getting polarimetric data into the hands of the largest number of researchers and students.

3. *Develop polarimetric O/IR synoptic and survey capabilities on intermediate-size telescopes to study YSOs/disks, probe their time evolution, and to promote student training in instrumentation and polarimetric observations*.
   - To be able to compete scientifically, we must invest in the next generation of young astronomers who will use polarimetry as a powerful tool in their light analysis toolbox

---

[2] See the All-Sky Polarimetry Survey White Paper (on http://astroweb.iag.usp.br/~mario/) submitted to the LSST Consortium

and who will understand polarimetric light analysis well enough to guide future instrument development.

- Wide-field polarimetric surveys, especially in the infrared, are needed to probe into extincted regions of star formation to discover and measure circumstellar and protostellar environments.
- Synoptic polarimetric observational data sets are crucial to understanding phenomena with complex geometry and/or time evolution.
- Obtaining ground-based calibrating polarimetric observations are crucial to the calibration of existing and future space-based polarimeters.

## *Final Thought*

The U.S. astronomical community has lost opportunities to advance key science areas as a result of down-selects of instrument capabilities or lack of will to commission polarimetric modes on instruments. The investment is minor, the expertise is available in the community, and the rewards are tangible. We are excited by the recent momentum favoring polarimetric studies and capabilities and believe the upcoming decade will see the various polarimetric techniques together become a strong, necessary component of astronomers' light analysis toolbox.

## *Bibliography and References*

Alves, F. O., Franco, G. A. P., & Girart, J. M. 2008, "Optical polarimetry toward the Pipe nebula: revealing the importance of the magnetic field," A&A, 486, L13

Andersson, B-G, & Potter, S. B. 2005, "A high sampling-density polarization study of the Southern Coalsack," MNRAS, 356, 1088

Andersson, B-G, & Potter, S. B. 2007, "Observational Constraints on Interstellar Grain Alignment," ApJ, 665, 369

Benjamin, R. A., et al. 2003, "GLIMPSE. I. An SIRTF Legacy Project to Map the Inner Galaxy," PASP, 115, 953

Bowen, D.V. et al., 2008, "The Far Ultraviolet Spectroscopic Explorer Survey of O VI Absorption in the Disk of the Milky Way," ApJS, 176, 59

Brown, J.C., et al. 2007, "Rotation Measures of Extragalactic Sources behind the Southern Galactic Plane: New Insights into the Large-Scale Magnetic Field of the Inner Milky Way", ApJ, 663, 258

Chandrasekhar, S., & Fermi, E. 1953, "Magnetic Fields in Spiral Arms," ApJ, 118, 113

Clemens, D. P., Sarcia, D., Grabau, A., Tollestrup, E. V., Buie, M. W., Dunham, E., & Taylor, B. 2007, "Mimir: A Near-Infrared Wide-Field Imager, Spectrometer and Polarimeter," PASP, 119, 1385

Crutcher, R. M., Hakobian, N., & Troland, T. H. 2008, "Response to the Mouschovias-Tassis Comments on "Testing Magnetic Star Formation Theory," arXiv0808.1150

Draine, B. T., & Lazarian, A. 1998, "Electric Dipole Radiation from Spinning Dust Grains," ApJ, 508, 157

Duffin, D. F., & Pudritz, R. E. 2008, "Simulating hydromagnetic processes in star formation: introducing ambipolar diffusion into an adaptive mesh refinement code," MNRAS, 391, 1659

Elmegreen, B. G. 2007, "On the Rapid Collapse and Evolution of Molecular Clouds," ApJ, 668, 1064

Falceta-Gonçalves, D., Lazarian, A., & Kowal, G. 2008, "Studies of Regular and Random Magnetic Fields in the ISM: Statistics of Polarization Vectors and the Chandrasekhar-Fermi Technique," ApJ, 679, 537

Galli, D., Shu, F. H., Laughlin, G., & Lizano, S. 2001, "Singular Isothermal Disks. II. Nonaxisymmetric Bifurcations and Equilibria," ApJ, 551, 367

Gnedin, N. Y., Ferrara, A., & Zweibel, E. G. 2000, "Generation of the Primordial Magnetic Fields during Cosmological Reionization," ApJ, 539, 505

Elmegreen, B. G. 2007, "On the Rapid Collapse and Evolution of Molecular Clouds," ApJ, 668, 1064

Han, J.L., Manchester, R.N., Lyne, A.G., Qiao, G.J. & van Straten, W., 2006, Pulsar Rotation Measures and the Large-Scale Structure of the Galactic Magnetic Field", ApJ, 642, 868

Harries, T. J. 2000, "Synthetic line profiles of rotationally distorted hot-star winds," MNRAS, 315, 722

Heiles, C. 2000, "9286 Stars: An Agglomeration of Stellar Polarization Catalogs," AJ, 119, 923

Hoang, T., & Lazarian, A. 2008, "Radiative torque alignment: essential physical processes," MNRAS, 388, 117

Hoeflich, P. 2005, "Radiation hydrodynamics in supernovae," Ap&SS, 298, 87

Hoffman, J. L. 2007, "Supernova polarization and the Type IIn classification," in AIP Conf. Ser. 937 (Supernova 1987A: 20 Years After: Supernovae and Gamma-Ray Bursters), eds. S. Immler, K.W. Weiler, & R. McCray (New York: AIP), 365

Hough, J. H., Aitken, D. K., Whittet, D. C. B., Adamson, A. J., Chrysostomou, A. 2008, "Grain alignment in dense interstellar environments: spectropolarimetry of the 4.67-µm CO-ice feature in the field star Elias 16 (Taurus dark cloud)," MNRAS, 387, 797

Jackson, J. M., et al. 2006, "The Boston University-Five College Radio Astronomy Observatory Galactic Ring Survey," ApJS, 163, 145

Joos, F. 2007, "Polarimetric direct detection of extra-solar planets with SPHERE/ZIMPOL," Proceedings of the conference "In the Spirit of Bernard Lyot: The Direct Detection of Planets and Circumstellar Disks in the 21st Century," lyot.confE, 28

Kasen, D., Thomas, R., & Nugent, P. 2006, "Time dependent Monte Carlo radiative transfer calculations for 3-dimensional supernova spectra, lightcurves, and polarization," ApJ, 651, 366

Kudoh, T., Basu, S., Ogata, Y., & Yabe, T. 2007, "Three-dimensional simulations of molecular cloud fragmentation regulated by magnetic fields and ambipolar diffusion," MNRAS, 380, 499

Landi degl'Innocenti, E. 1999, "Evidence for ground-level atomic polarization in the solar atmosphere," ASSL, 243, 61

Lazarian, A., & Hoang, T. 2007a, "Radiative torques: analytical model and basic properties," MNRAS, 378, 910

Lazarian, A., & Hoang, T. 2007b, "Subsonic Mechanical Alignment of Irregular Grains," ApJ, 669, L77

Leroy, J.P., 1999, "Interstellar dust and magnetic field at the boundaries of the Local Bubble. Analysis of polarimetric data in the light of HIPPARCOS parallaxes", A&A, 346, 955

Li, P. S., Norman, M. L., Mac Low, M.-M., & Heitsch, F. 2004, "The Formation of Self-Gravitating Cores in Turbulent Magnetized Clouds," ApJ, 605, 800

Macintosh, B., et al. 2006, "The Gemini Planet Imager," SPIE, 6272, 18

Nordsieck, K., 2008, "Atomic Fluorescence and Prospects for Observing Magnetic Geometry Using Atomic Magnetic Realignment", arXiv:0809.3802

Ostriker, E. C., Stone, J. M., & Gammie, C. F. 2001, "Density, Velocity, and Magnetic Field Structure in Turbulent Molecular Cloud Models," ApJ, 546, 980

Padoan, P., & Nordlund, Å. 2002, "The Stellar Initial Mass Function from Turbulent Fragmentation," ApJ, 576, 870

Reynolds, R.J., Haffner, L. M., Madsen, G. J., & Tufte, S. L., 2004, "Surveying the Galaxy with WHAM", ASPC, 317, 186

Stil, J. M., 2006, "The VLA Galactic Plane Survey", AJ, 132, 1158

Tamura, M., et al. 2007, "Near-Infrared Imaging Polarimetry of the NGC 2071 Star-Forming Region with SIRPOL," PASJ, 59, 467

Tassis, K., & Mouschovias, T. Ch. 2007, "Protostar Formation in Magnetic Molecular Clouds beyond Ion Detachment. I. Formulation of the Problem and Method of Solution," ApJ, 660, 370

Whittet, D.C.B., Adamson, A.J., Duley, W.W., Geballe, T.R. & McFadzean, A.D., 1989, "Infrared spectroscopy of dust in the Taurus dark clouds - Solid carbon monoxide", MNRAS, 241, 707.

Whittet, D. C. B., Hough, J. H., Lazarian, A., & Hoang, T. 2008, "The Efficiency of Grain Alignment in Dense Interstellar Clouds: a Reassessment of Constraints from Near-Infrared Polarization," ApJ, 674, 304

Yan, H., & Lazarian, A. 2008, "Atomic Alignment and Diagnostics of Magnetic Fields in Diffuse Media", ApJ, 677, 1401.

Zweibel, E. G. 2003, "Cosmic-Ray History and Its Implications for Galactic Magnetic Fields," ApJ, 587, 625